\documentclass[a4paper]{easychair}

\RequirePackage{tabulary}
\usepackage{doc}
\usepackage{xspace}
\usepackage{textcomp}
\usepackage{booktabs}
\usepackage{scalerel}
\usepackage{tikz}
\usepackage{pgf-umlsd}
\usepackage{listings}
\usetikzlibrary{svg.path}
\usepackage{cite}
\usepackage{subcaption}
\usepackage{xcolor}
\usepackage{url}

\title{Coupling Microscopic Mobility and Mobile Network Emulation
for Pedestrian Communication Applications
\thanks{We thank the research office (FORWIN) of the Munich University of Applied Sciences for supporting
the research collaboration. The authors gratefully acknowledge the support by the
Faculty Graduate Center \mbox{CeDoSIA} of TUM Graduate School at Technical University of Munich, Germany.
The authors also acknowledge the financial support by the Federal Ministry of Education and Research
of Germany in the framework of roVer (project number 13FH669IX6).
}
}

\author{
Matthias Rupp
\and
Stefan Schuhbäck
\and
Lars Wischhof
}

\institute{
  Dept. of Computer Science and Mathematics \\
  Munich University of Applied Sciences\\
  Munich, Germany\\
  \email{mrupp@hm.edu}, \email{stefan.schuhbaeck@hm.edu}, \email{lars.wischhof@hm.edu}
 }

\authorrunning{Rupp, Schuhbäck and Wischhof}

\titlerunning{Coupling Microscopic Mobility and Mobile Network Emulation}

\begin{document} 

\maketitle

\newcommand{\opp}{\mbox{OMNeT++}\xspace}
\newcommand{\vadere}{Vadere\xspace}
\newcommand{\inet}{INET\xspace}
\newcommand{\veins}{Veins\xspace}
\newcommand{\sumo}{SUMO\xspace}
\newcommand{\simulte}{SimuLTE\xspace}
\newcommand{\pedemap}{PeDeMaP\xspace}
\newcommand{\crownet}{CrowNet\xspace}

\begin{abstract}
  Network emulation is a well-established method for demonstrating and testing
  real devices and mobile apps in a controlled scenario.
  This paper reports preliminary results for an open-source extension of the \crownet 
  pedestrian communication framework. It enables the interaction between simulated and 
  real devices using the emulation feature of \opp. The interaction is handled by several 
  \opp modules that can be combined to match different use-cases. 
  Initial timing measurements have been conducted for an example application which creates decentralized
  pedestrian density maps based on pedestrian communication. 
  The results indicate that the approach is feasible for scenarios with a limited number of pedestrians.
  This limitation is mainly due to the real-time simulation requirements in coupled emulation.
\end{abstract}

\section{Introduction}
\label{sect:introduction}
In Intelligent Transportation Systems (ITS) and related areas, wireless communication of
vehicles and infrastructure is of fundamental importance and has been a major focus
of research for more than two decades. Vehicular communication protocols and suitable
communication models and simulation tools have been developed and are widely used.
Communication of and with pedestrians, 
despite often being mentioned as a possible "X" in Vehicle-to-X (V2X) communication, 
did not receive that much attention.
This changed in recent years, when the potential of pedestrian communication
for safety and protection of human lives received more attention, e.g. suitable
message formats such as the Vulnerable Road User (VRU) Awareness Messages 
(VAM)~\cite{etsi-2020-TS1033003-com} were standardized and are considered
in industry forums such as the 5G Automotive Association (5GAA). In parallel, also the research in this
area gained momentum, e.g. to evaluate the impact of VAM generation rate adaptation 
on the awareness of VRUs \cite{lara-2021-com}.

When these techniques are deployed in future, it is very likely that the smartphone
of a pedestrian will send these and other messages, i.e. it will be the
main communication device of a pedestrian -- comparable to the role that an 
OnBoard Unit (OBU) has for vehicles. Thus, the research community needs to
develop mobile apps for pedestrian communication, test them in realistic settings and
must be able to demonstrate the benefits of the applications. As a first step to enable the two latter
ones, this paper presents a straight-forward extension of the \opp based
open-source framework \crownet, which allows to run a pedestrian communication
app on an Android smartphone within an emulated pedestrian communication
scenario. \crownet itself combines several public \opp models (see Sec.~\ref{arch:coupling}) to 
simulate the interaction between mobile communication and pedestrian mobility.

Our main motivations for developing the extension are threefold: a) testing mobile 
applications in a reproducible pedestrian communication scenario, e.g. within a Continuous Integration (CI) pipeline,
b) demonstrating the usefulness of a mobile app based on pedestrian communication,
c) enabling user-studies. While we assume that the framework is suitable for
a wide range of applications, we illustrate its usage on the example of
a mobile app generating pedestrian density maps.

\textbf{Contributions:} The main contributions of this paper are the following:
1. a novel open-source extension of the CrowNet \opp framework which allows 
  to test and demonstrate pedestrian communication based apps on real smartphones,
  2. illustrating the concept using an example-application which generates local 
  pedestrian density maps,
  3. preliminary results proving the feasibility of the proposed approach.

\section{Related Work}
\label{sect:relatedwork}
While we are not aware of any other public open-source framework for
network emulation specifically for mobile apps for pedestrians, related
work exists regarding several aspects.

\subsection{Network Emulation}
Emulation test beds for mobile apps have been investigated by several 
research projects: Hetu et.al.~developed a tool called Similitude \cite{hetu-2014-com}, which
couples the road traffic simulator \sumo with several Android mobile device
emulators and (optionally) the open-source network simulator ns-3 
to evaluate intelligent transportation systems. Although
their concept of coupling Android devices and network simulation is somewhat similar
to our approach, the CrowNet emulation presented here is designed for
pedestrian communication and therefore includes specific pedestrian mobility models 
(see Sec.~\ref{relwork:mobility} and \ref{arch:coupling}). A different approach for
testing Android applications was presented in \cite{serban-2015-com}, where a 
hybrid emulation test bed specifically for military applications was presented. However,
it does neither include a detailed model for 4G/5G cellular communication nor a 
detailed pedestrian mobility model.

The general aspect of testing real devices and applications within test beds 
applying network emulation has also been considered multiple authors: 
E.g., in \cite{sakai-2015-com}, Saki et. al. use ns-3 with the LENA LTE model 
to perform system-level network emulation of a LTE network and demonstrate that it can be used to predict 
the performance of VoIP applications in real LTE networks.
More recently, Nardini et. al. published results on using the system-level network simulator 
Simu5G in an emulation test bed
for multi-access edge computing \cite{nardini-2020a-com}. 

\subsection{Pedestrian Mobility Simulation}
\label{relwork:mobility}

A realistic modelling of the pedestrian mobility has a significant influence on
the observed properties of a mobile network \cite{hahn-2015-com}. 
While simplified mobility models for pedestrians are available in most traffic simulation systems,
an accurate microscopic modelling requires specific models such as the 
Social Force Model (SFM) \cite{farina-2017-cdyn} 
or the more advanced Optimal Steps Model (OSM) \cite{sivers-2015d-cdyn, kleinmeier-2019-cdyn}
Therefore, several pedestrian and crowd simulation frameworks
have been implemented. The open-source crowd simulator Vadere\cite{kleinmeier-2019-cdyn} 
is widely used and has been validated by several 
experimental studies. It is used for pedestrian mobility simulation in this paper.




\section{System Architecture}


\begin{figure*}
  \centering
  \begin{subfigure}[t]{0.78\textwidth}
    \centerline{\includegraphics[width=\textwidth]{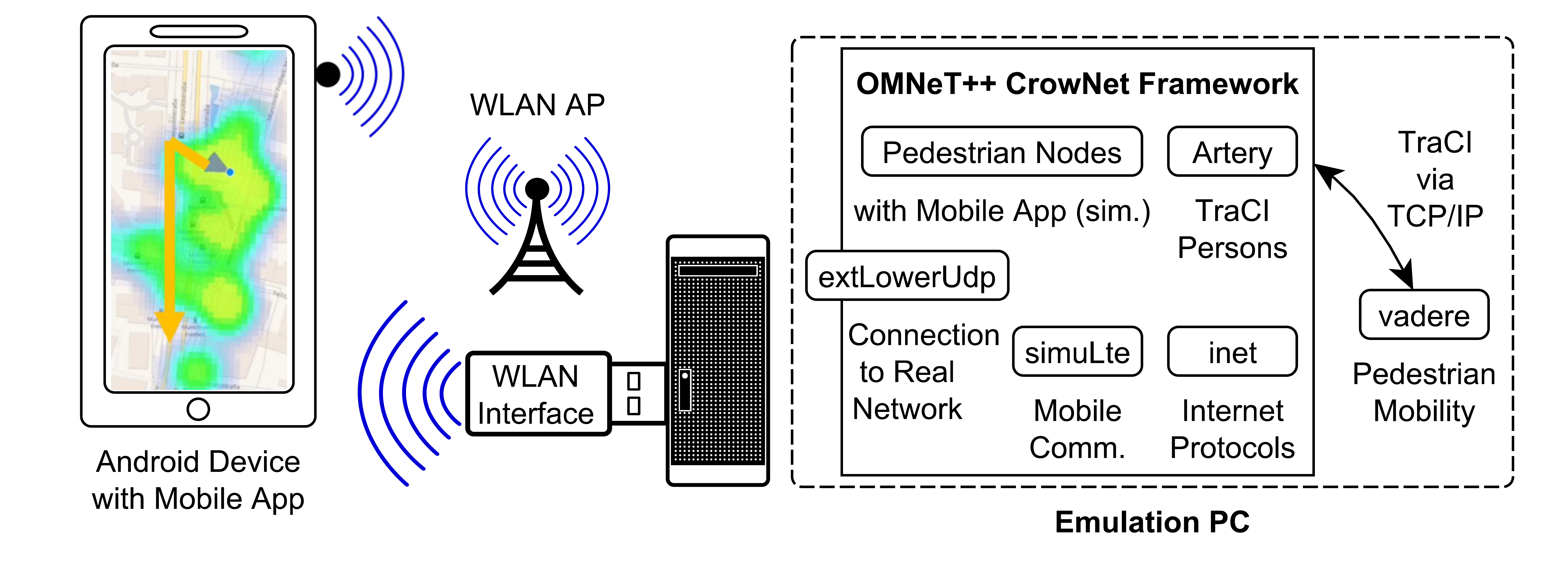}}
    \caption{Simplified Overview}
    \label{fig:emulation_structure}
  \end{subfigure}%
  ~ 
  \begin{subfigure}[t]{0.2\textwidth}
    \centering
    \includegraphics[width=\textwidth]{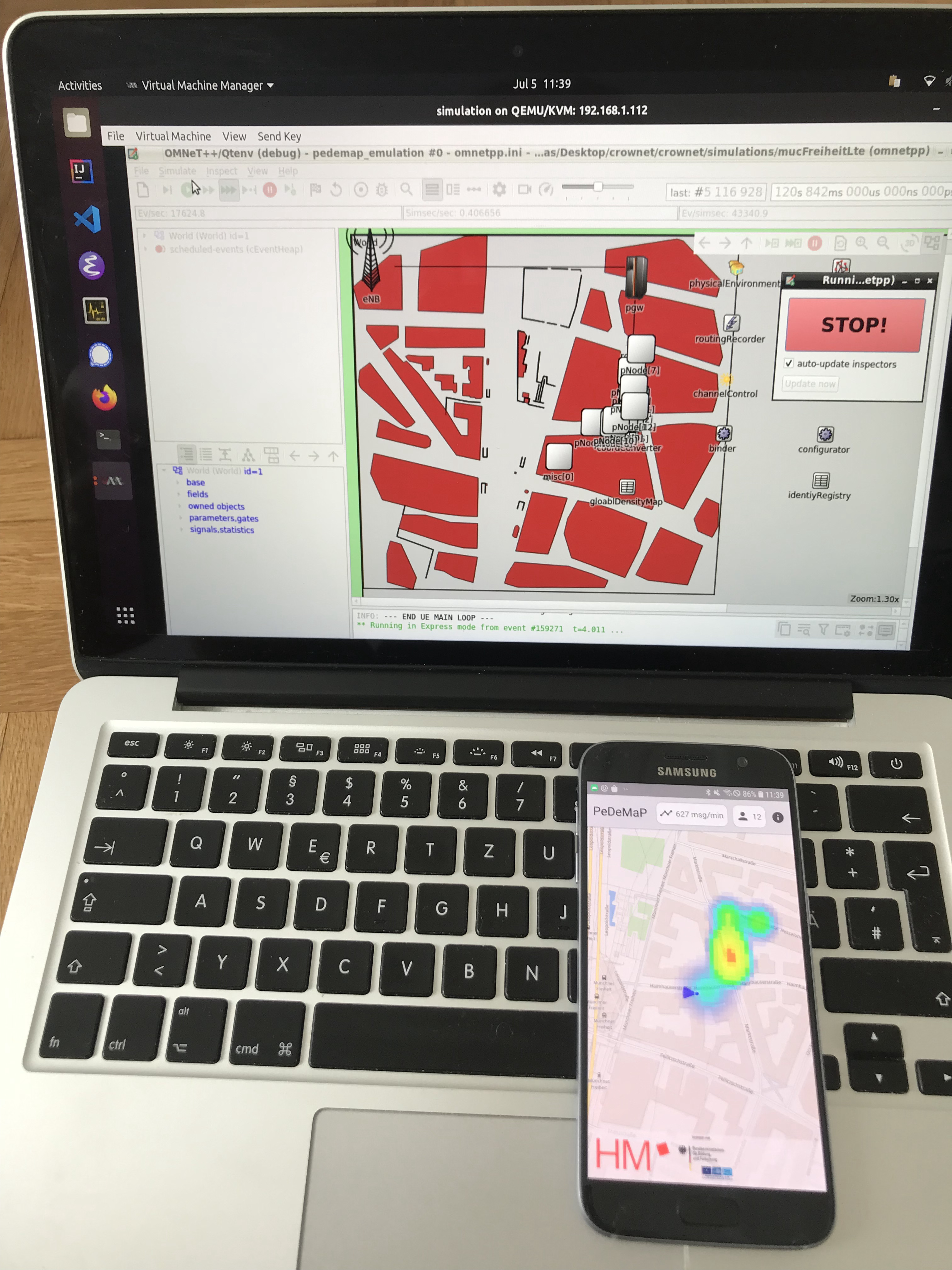}
    \caption{Lab Setup}
    \label{fig:android_sim}
  \end{subfigure}
  \caption{System architecture: Emulation environment for pedestrian communication.}
\end{figure*}

The basic idea is illustrated in Fig.~\ref{fig:emulation_structure}: 
One or more mobile devices are connected via local WLAN to a PC which 
executes the emulation environment.\footnote{We assume that the data rate 
available on the local WLAN link is high and that the additional
delay is low compared to the delay caused by pedestrian communication and 
that it can therefore be neglected.} Packets sent 
by the real mobile device are captured by the network interface of the
emulation PC and injected into the emulation environment (via the
\emph{extLowerUdp} module, as explained in more detail in 
Sec.~\ref{sec:network_emulation_mobile_apps}). Packet injection is performed by
a place-holder of the real device within the simulation.  Thus, packets are
received and processed by other (simulated) pedestrians within communication
range.  Vice versa, packets received by the real-device's place-holder are sent
via WLAN to the real mobile device and processed there.  Since a realistic
mobility simulation is of high importance, we use \vadere 
(see Sec.~\ref{sect:relatedwork}) as a mobility provider in the coupled emulation.

The next subsections first present the basic approach for coupling pedestrian mobility
simulation with mobile network simulation. Afterwards, an example application is outlined.

\subsection{Coupling Pedestrian Mobility and Mobile Network Simulation}
\label{arch:coupling}

CrowNet\footnote{Sources are publicly available at 
\url{https://github.com/roVer-HM/crownet/tree/emulation\_omnetsummit21},
general information on the CrowNet framework can be found at 
\texttt{https://crownet.org}} is an open-source simulation framework 
that combines multiple simulators from the
communication and mobility domain. It allows the simulation of scenarios where
the interaction between these domains is an integral part of the simulation
study. For the communication domain, the discrete event simulator OMNeT++ (version 6.0 Preview 11) is used
in conjunction with the widely-used models
INET (version 4.3.2), 
Artery~\cite{riebl-2015-com}, SimuLTE~\cite{nardini-2018-com}/Simu5G~\cite{nardini-2020a-com}. 
For the mobility domain, \sumo~\cite{lopez-2018-com} as well as the crowd dynamics framework
Vadere~\cite{kleinmeier-2019-cdyn} (see Sec.~\ref{sect:relatedwork}) are supported.

To connect the simulators, CrowNet uses the Traffic Control Interface (TraCI) as the
inter-process communication to interlock the simulation loops of the
selected simulators. This synchronization allows the exchange and modification of 
simulation state between the mobility and communication domain.  The initial
integration (\cite{schuhbaeck-2019-com}) was based on Veins\cite{sommer-2011-com}. 
However, the current TraCI connection management uses
Artery to harmonize the implementation between Vadere and \sumo nodes.

The example scenario presented in this paper uses broadcast LTE-A side link
communication (PC5 interface) to disseminate pedestrian density maps. 
For modelling pedestrian mobility, the Optimal
Steps Model (OSM), provided by \vadere, is applied. The OSM model uses navigation
fields  which encode the shortest geodesic distance from the current position
to a pre-selected target while modelling each step \cite{kleinmeier-2019-cdyn}. 

\subsection{Use-Case Example: Decentralized Pedestrian Density Maps}

Emulation based test beds are helpful to test real application 
in network situation which are not easy to reproduce in real life. 
As an example, we use a decentralized application that 
provides pedestrian density information for the local area in the form of a  Decentralized 
Pedestrian Density (DPD) map. 

\begin{figure}[htbp]
  \centerline{\includegraphics[width=0.9\textwidth]{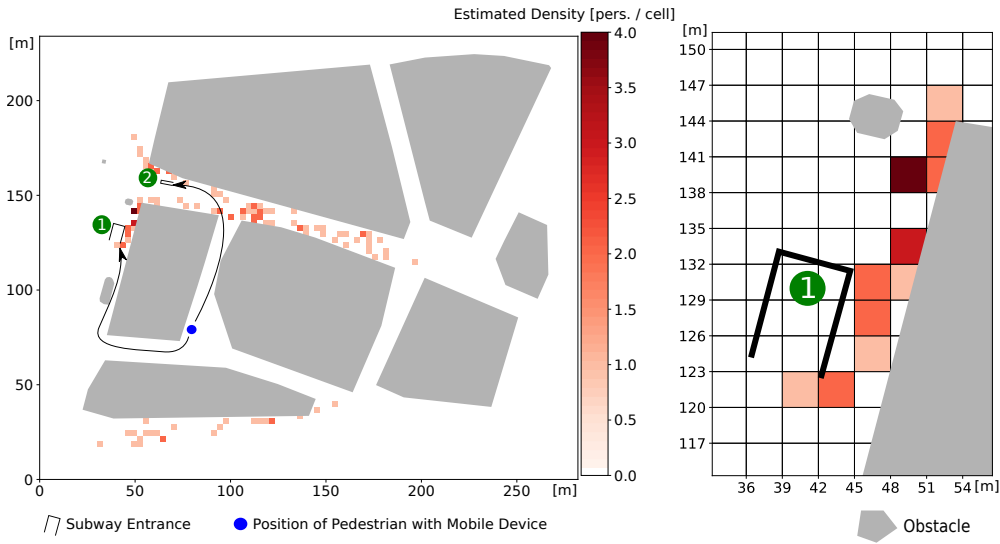}}
      \caption{Left: DPD map (based on \cite{schuhbaeck-2021-com}).
      The highlighted pedestrian (blue dot) can select an alternative route.
      Right: Enlarged view of target area. Color indicates estimated density.}
  \label{fig:map}
  \end{figure}

DPD maps allow the creation of decentralized awareness of the surrounding area 
which is an important context parameter needed for multiple urban centric mobility 
applications such as high crowed density detection and avoidance (e.g. to reduce 
Covid-19 infections), route and capacity planing for public transportation 
or individual real time route planing.

DPD maps are created by 
aggregating  periodically broadcasted position beacons from neighboring nodes 
into a two dimensional density map of the surrounding area (Fig.~\ref{fig:map}). 
These maps are  re-broadcasted to share the local density data with neighboring nodes, allowing them 
to merge this view with there own perception and create a larger area
of awareness~\cite{schuhbaeck-2021-com}. 

The decentralized creation and dissemination approach makes it hard to test the
application on real hardware. To get meaningful scenarios, multiple
devices have to be used by real pedestrians in a real urban setup. These tests
are time and resource expensive. 
On the other hand, only testing the scenario in a simulated environment does not
take into account the effects of real hardware. 
Thus, using real devices within the \opp emulation testbed allows to have the 
advantages of real devices but still achieves reproducibility and scalability though simulation.

\subsubsection{DPD Android App}
While the simulation model for DPD maps example was already publicly available 
\cite{schuhbaeck-2021-com}, for testing the proposed emulation framework, an
identical application had to be implemented on a real Android smartphone.
It was implemented in Kotlin using the usual Android development tools.
This Android app calculates and visualizes the pedestrian density in the surrounding area. 
This is achieved by exchanging location messages, serialized as Protocol Buffers,  
between participating devices.

For data dissemination in a local area, the Android application currently cannot use
LTE-A sidelink communication via PC5 interface, since the available commercial
smartphones do not implement this mode of cellular communication, yet. Therefore, the
Android app uses broadcast messages sent via WLAN for local communication as a
workaround. 

Density maps consist of rectangular cells, e.g. $3\text m \times 3\text m$. 
The real device obtains its local position using the Android Location Services in 
form of WGS84 coordinates. In order to simplify the mapping of the current position
to a DPD cell, a Universal Transverse Mercator (UTM) projection is applied.
Each cell then is identified by the UTM coordinates of its top-left corner.


When two real devices communicate, the WGS84-based location is converted to UTM, encoded as
DPD location data and sent in form of a LocationMessage via UDP/IP/WLAN broadcast to the second
device. 




\section{Network Emulation for Mobile Apps}
\label{sec:network_emulation_mobile_apps}


\begin{figure}[tb]
  \centering
  \begin{subfigure}[t]{0.3\textwidth}
    \centering
    \resizebox{5cm}{3.4cm}{
    \begin{tikzpicture}
      \draw [dashed] (0, 3.3) rectangle (5.5, 4.9);

        \node [draw, label=below:\scriptsize{\texttt{BeaconApp}}, minimum width=0.7cm, minimum height=0.7cm] (ba) at (1.3, 4.3){};
        \node [draw, label=below:\scriptsize{\texttt{DensityMapApp}}, minimum width=0.7cm, minimum height=0.7cm] (dm) at (3.7, 4.3){};

      \draw [latex-latex] (1.3, 3.3) -- (ba);
      \draw [latex-latex] (3.7, 3.3) -- (dm);

      \draw [dotted] (0, 2.2) rectangle (5.5, 3.2);
      \node (3) at (1.15, 3.0) {\small{\emph{transport layer}}};
        \node [draw, label=right:\small{\texttt{udp}}, minimum width=0.7cm, minimum height=0.2cm] (udp) at (3.5, 2.7){};
      \draw [latex-latex] (3.5, 2.2) -- (udp);
      \draw [latex-latex] (3.5, 3.2) -- (udp);

      \draw [dotted] (0, 1.1) rectangle (5.5, 2.1);
      \node (2) at (1.05, 1.9) {\small{\emph{network layer}}};
      \node [draw, label=right:\small{\texttt{ipv4}}, minimum width=0.7cm, minimum height=0.2cm] (ip) at (3.5, 1.6){};
    \draw [latex-latex] (3.5, 1.1) -- (ip);
    \draw [latex-latex] (3.5, 2.1) -- (ip);

    \draw [dotted] (0, 0) rectangle (5.5, 1.0);
    \node (1) at (0.8, 0.8) {\small{\emph{link layer}}};
    \node [draw, label=right:\small{\texttt{lteNic}}, minimum width=0.7cm, minimum height=0.2cm] (ln) at (3.5, 0.5){};
  \draw [latex-latex] (3.5, 0.0) -- (ln);
  \draw [latex-latex] (3.5, 1.0) -- (ln);
    \end{tikzpicture}
    }
    \caption{\texttt{node[*]}}
    \label{fig:nodes:1}
  \end{subfigure}
  \begin{subfigure}[t]{0.65\textwidth}
    \centering
    \resizebox{6cm}{4.4cm}{
    \begin{tikzpicture}
      
      \draw [dashed] (0, 3.3) rectangle (6.5, 6.3);

        \node [draw, label=below:\scriptsize{\texttt{OutboundEmulation}}, minimum width=0.7cm, minimum height=0.7cm] (ob) at (1.5, 4.2){};
      \node [draw, label=below:\scriptsize{\texttt{inet::ExtLowerUdp}}, minimum width=0.7cm, minimum height=0.7cm] (elu) at (1.5, 5.7){};
      \node [draw, label=below:\scriptsize{\texttt{NodeLocationExporter}}, minimum width=0.7cm, minimum height=0.7cm] (nle) at (4.5, 4.2){};
      \node [draw, label=below:\scriptsize{\texttt{inet::ExtLowerUdp}}, minimum width=0.7cm, minimum height=0.7cm] (elu2) at (4.5, 5.7){};

      
      \draw [latex-latex] (1.5, 3.3) -- (ob);
      \draw [latex-latex] (ob) -- (elu);
      \draw [latex-latex] (nle) -- (elu2);

      \draw [dotted] (0, 2.2) rectangle (6.5, 3.2);
      \node (3) at (1.15, 3.0) {\small{\emph{transport layer}}};
        \node [draw, label=right:\small{\texttt{udp}}, minimum width=0.7cm, minimum height=0.2cm] (udp) at (3.5, 2.7){};
      \draw [latex-latex] (3.5, 2.2) -- (udp);
      \draw [latex-latex] (3.5, 3.2) -- (udp);

      \draw [dotted] (0, 1.1) rectangle (6.5, 2.1);
      \node (2) at (1.05, 1.9) {\small{\emph{network layer}}};
      \node [draw, label=right:\small{\texttt{ipv4}}, minimum width=0.7cm, minimum height=0.2cm] (ip) at (3.5, 1.6){};
    \draw [latex-latex] (3.5, 1.1) -- (ip);
    \draw [latex-latex] (3.5, 2.1) -- (ip);

    \draw [dotted] (0, 0) rectangle (6.5, 1.0);
    \node (1) at (0.8, 0.8) {\small{\emph{link layer}}};
    \node [draw, label=right:\small{\texttt{lteNic}}, minimum width=0.7cm, minimum height=0.2cm] (ln) at (3.5, 0.5){};
  \draw [latex-latex] (3.5, 0.0) -- (ln);
  \draw [latex-latex] (3.5, 1.0) -- (ln);
      
    \end{tikzpicture}
    }
    \caption{\texttt{node[0]}}
    \label{fig:nodes:0}
  \end{subfigure}
  \caption{Structure of a generic node and \texttt{node[0]}, serving as emulation bridge. The position of \texttt{node[0]} is controlled by \vadere, therefore, the InboundEmulation module is not needed.}
  \label{fig:nodes}
\end{figure}
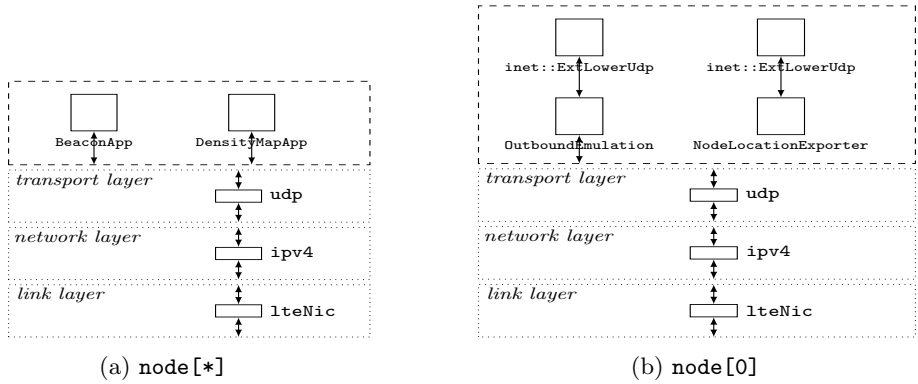

\begin{figure}[tbh]
  \centering

  \begin{subfigure}[t]{0.27\textwidth}
    \centering
    \begin{tikzpicture}
      \node[draw] (p) at (1, 2) {App};
      \node[draw] (n)  at (1, 1) {\texttt{node[0]}};
      \node[draw] (v) at (0, 0) {\vadere};
      \node[draw] (nn) at (2, 0) {\texttt{node[*]}};

      \draw[-latex, linkcolor, line width=0.7] (v) -- (nn);
      \draw[-latex, linkcolor, line width=0.7] (v) -- (n);
      \draw[-latex, linkcolor, line width=0.7] (n) -- (p);
    \end{tikzpicture}
    \caption{NodeLocationExporter}
    \label{fig:posflow:b}
  \end{subfigure}
  \hspace{5em}
  \begin{subfigure}[t]{0.27\textwidth}
    \centering
    \begin{tikzpicture}
      \node[draw] (p) at (1, 2) {App};
      \node[draw] (n)  at (1, 1) {\texttt{node[0]}};
      \node[draw] (v) at (0, 0) {\vadere};
      \node[draw] (nn) at (2, 0) {\texttt{node[*]}};

      \draw[-latex, linkcolor, line width=0.7] (p) -- (n);
      \draw[-latex, linkcolor, line width=0.7] (v) -- (nn);
    \end{tikzpicture}
    \caption{InboundEmulation}
    \label{fig:posflow:c}
  \end{subfigure}
  \caption{Flow of mobility data between \vadere, the simulated nodes, and the coupled app.}
  \label{fig:posflow}
\end{figure}
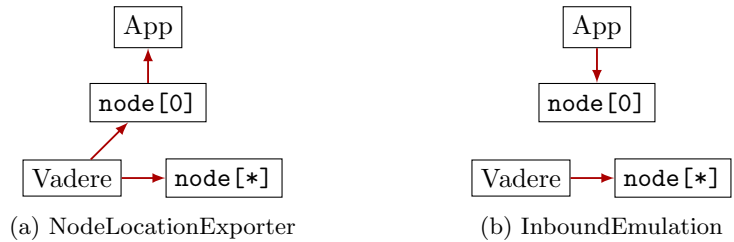

To meet our goal of connecting Android applications running in a real network to the simulation, we developed a bridge that forwards and converts the  position beacons between the real and emulated network. The offset between real- and simulated world coordinates can be specified in the modules' configurations. Each bridge-module connects one real device to the simulation.
The bridge-module is located in the application layer of a specific node, which now represents the real device inside the simulation. 
In scenarios where one device is connected, the bridging is handled by \texttt{node[0]} (Fig.~\ref{fig:nodes}).
The node does not need any applications for the dissemination of beacons and density maps because this task is handled by the coupled device.
The bridge-module consists of three modules:
\begin{description}
  \item[OutboundEmulation] receives the position beacons from \texttt{node[1..n]} and forwards them to the mobile application, 
  applying necessary coordinate projections. The module consists of 2 sub-modules: 
  The \texttt{DensityMessageHandler} receives position beacons and converts them to the required protocol buffers format 
  for the mobile application. \texttt{DensityMessageHandler} is connected to an \texttt{inet::ExtLowerUdp} module which 
  sends the data over the real network to the connected devices via an UDP socket.
  \item[NodeLocationExporter] exports the simulated position of \texttt{node[0]} to the mobile application.
  This allows to mock the device's location using mobility data from \vadere. Again, the \texttt{inet::ExtLowerUdp} 
  module is used to send the data to the real-device(s).
  \item[InboundEmulation] receives protocol buffers location beacons generated by the mobile application 
  and sets the position of \texttt{node[0]} to the position of the real device. 
  The beacons are received by a \texttt{inet::ExtLowerUdp} socket which is connected to the \texttt{InboundEmulation} module. 
  The data is serialized to a simple string-based format.
\end{description}
The framework currently supports two modes for setting the location of the real mobile devices:
\begin{enumerate}
\item If the NodeLocationExporter module is used (see Fig.~\ref{fig:posflow:b}), the app's location 
is controlled by \vadere. This allows to test location-aware applications without the need of 
actually moving the device which makes these tests 
reproducible and independent of error sources such as the GPS reception.If the NodeLocationExporter module is used (see Fig.~\ref{fig:posflow:b}), the app's location 
is controlled by \vadere. This allows to test location-aware applications without the need of 
actually moving the device which makes these tests 
reproducible and independent of error sources such as the GPS reception.
\item If the InboundEmulation module receives mobility information from the real mobile application, it moves 
its place-holder (e.g. \texttt{node[0]}) accordingly (Fig.~\ref{fig:posflow:c}). This variant is mainly attractive for interactive demonstrations where the real device
is moved within a real urban scenario, which is replicated in the coupled simulation.
\end{enumerate}

\subsection{Scheduling}

To satisfy the timing requirements of both the real and simulated nodes, it is important that the simulation time is synchronized with the wall-clock time. The \texttt{inet::RealTimeScheduler}, which is required by the \texttt{inet::ExtLowerUdp} module, performs this synchronization. However, we observed that due to the fact that the scheduler starts before the connection to \vadere is established, there is an initial delay between $t_{sim}$ and $t_{real}$. Therefore we created a custom scheduler class called \texttt{crownet::EmulationScheduler}, which inherits from \texttt{inet::RealTimeScheduler} and includes a method to synchronize the simulation time back to the wall-clock time as well as some methods for timing analysis.
During testing, it showed that the most important factor whether or not the simulation time can keep up with the wall-clock time is 
the number of simulated nodes (see Sec.~\ref{sec:rt}).

\section{Measurement Results}
\label{sec:rt}

As a proof-of-concept, an urban simulation scenario with one eNodeB 
and a varying number of UEs/nodes was configured. Tab.~\ref{tab:para}~\footnote{
    For full simulation setup see \url{https://github.com/roVer-HM/crownet/tree/emulation_omnetsummit21/crownet/simulations/emulation_omnetsummit21}} lists 
the most relevant parameters, the complete sources are available in
the CrowNet repository, see Sec.~\ref{arch:coupling}.
To measure the timing behavior depending on the numbers of simulated nodes, 
both the real-time and simulation time were recorded at a 10 ms interval. 
Then, the offset between real-time and simulation time $t_{real} - t_{sim}$ 
was calculated. All measurements were performed on a low-end virtual 
machine (2 CPU-cores, 12GB RAM, Ubuntu Linux).

\begin{figure}[tb]
  \begin{subfigure}[t]{0.5\textwidth}
    \centering
    \includegraphics[height=5cm]{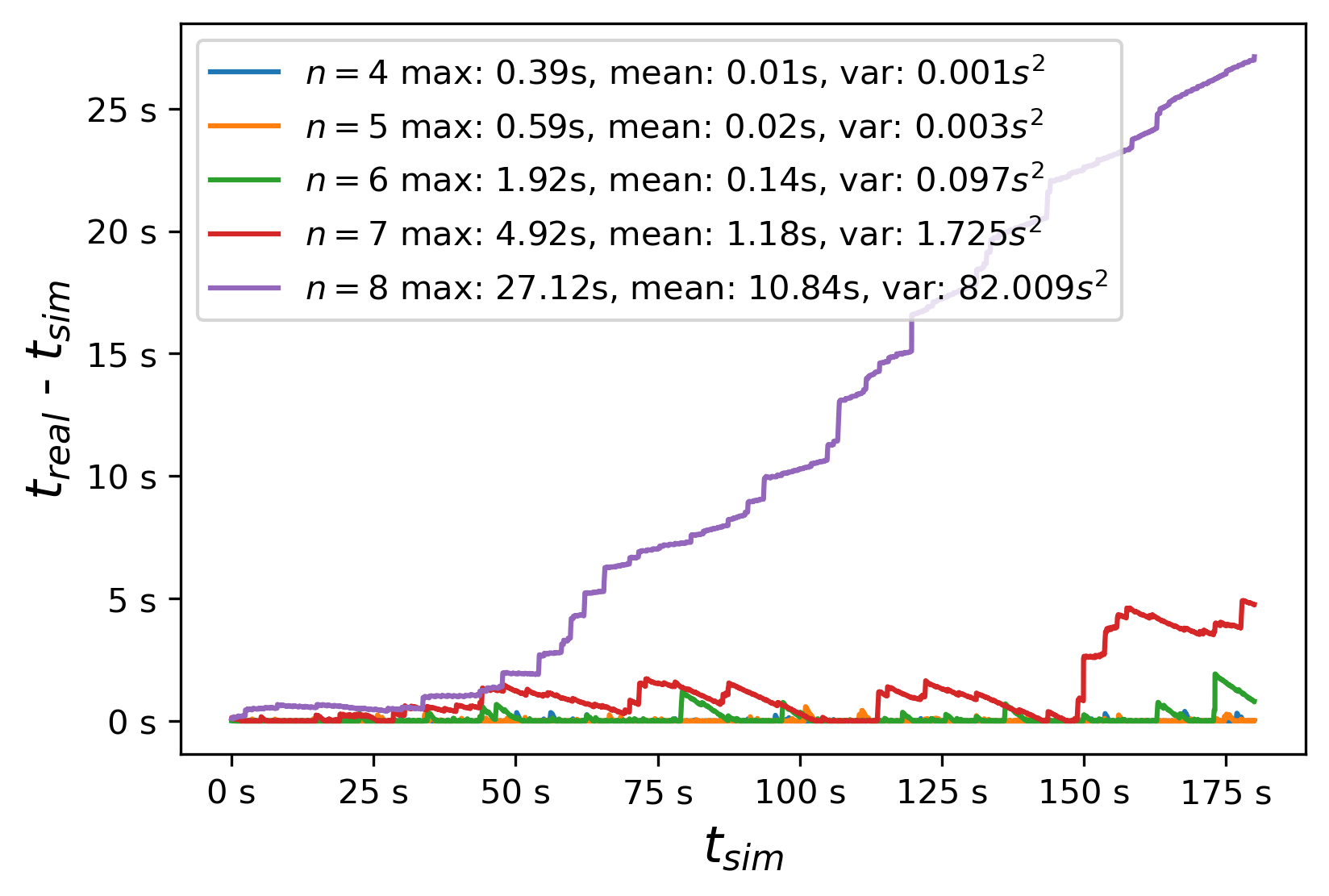}
    \caption{Overview}
    \label{fig:latency:full}
  \end{subfigure}
  \begin{subfigure}[t]{0.5\textwidth}
    \centering
    \includegraphics[height=5cm]{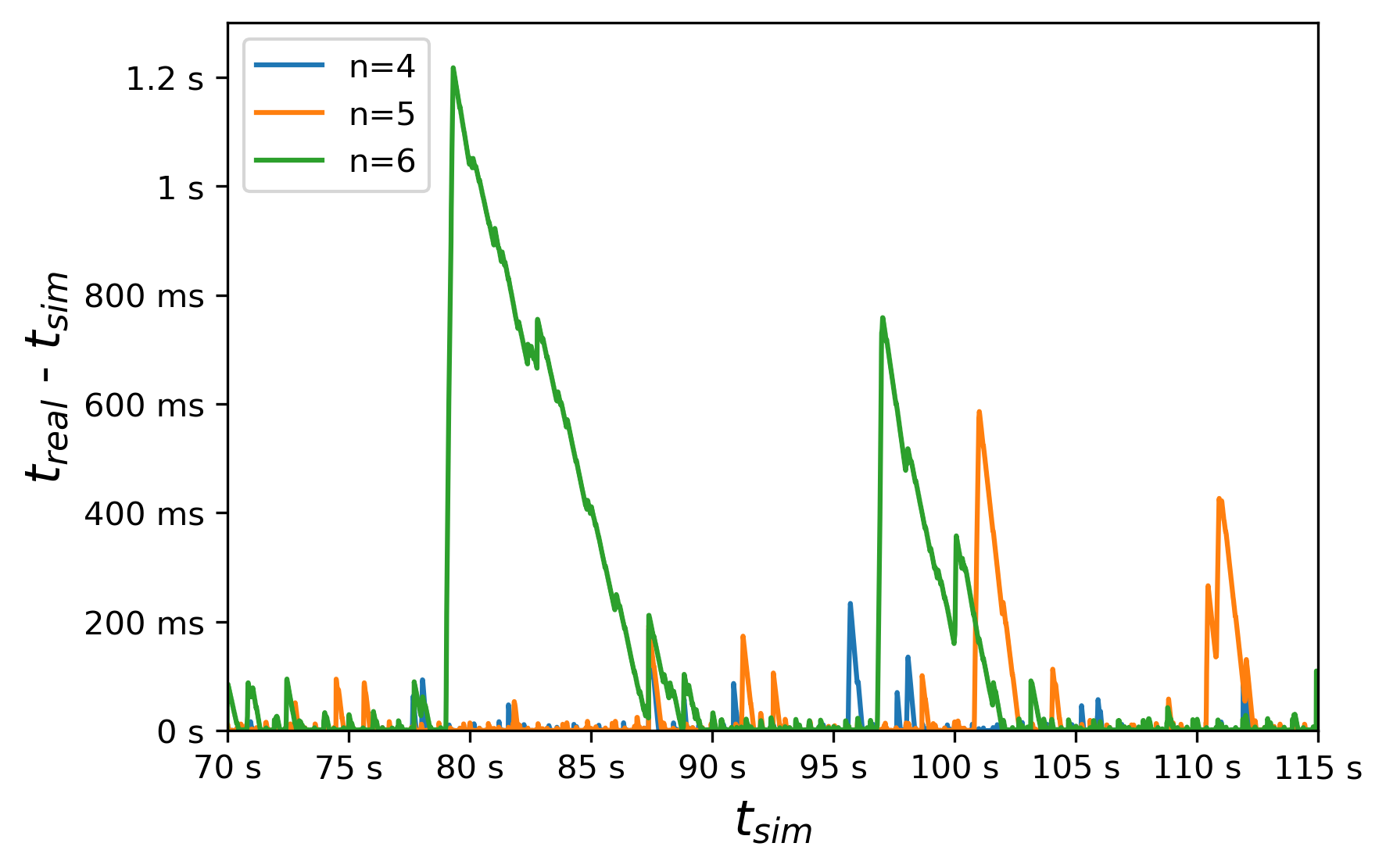}
    \caption{Detailed view of 80-115s for 4-6 nodes}
    \label{fig:latency:zoomed}
  \end{subfigure}
  \caption{Observed offset of real-time and simulation time ($t_{real} - t_{sim}$) depending 
  on the number of simulated nodes $n$. On a low-end machine, a value of  $n > 6$ leads to 
    offsets exceeding 5s.}
  \label{fig:latency}
\end{figure}

Fig.~\ref{fig:latency:full} shows the a plot for scenarios with 4 to 8 nodes.
Synchronized emulation is possible for up to 6 nodes on this machine. 
Fig.~\ref{fig:latency:zoomed} reveals that although scenarios with $n < 7$ in general
have a low offset $t_{real} - t_{sim}$ near zero, 
there are periods where the simulation time lags behind the real-time for several seconds. 
The lag increases with increasing number of nodes. However, the observed lags 
of $t_{sim} - t_{real} < 5s$ are negligible for our use-case.

\section{Conclusions}
The presented extension of the \crownet framework allows network emulation for mobile apps based on
pedestrian communication. However, emulation currently is only feasible for scenarios with
a very limited number of simulated nodes. For a low-end machine, the offset between real time and
simulation time can only be kept low for scenarios with up to six nodes.
Using our emulation strategy with larger scenarios would 
require a performance optimization of the simulation, for example a lower level of detail 
for the system-level mobile communication simulation or parallel execution. Despite the 
limited scalability, the presented framework is a useful tool for small-scale tests or
demonstrations with real hardware devices.

In future, we will upgrade the communication model to 5G (using Simu5G) and perform
in-depth measurements to analyze the system in detail. We will especially try to find the main cause of the performance problems using a profiler tool. The goal is also to 
integrate more data formats in our emulation framework to support more types of 
mobile applications.

\begin{table}[!htb]
    \caption{Most relevant parameters used for the emulation Config: pedmap\_emulation.}
  \begin{subtable}{.4\linewidth}
    \centering
      \caption{Pedestrian Mobility}
      \label{tab:para}
      \tiny
      \centering
      \settowidth\tymin{\textbf{Symbol}}
      \setlength\extrarowheight{3pt}
      \begin{tabulary}{0.95\linewidth}{LLL}
          \toprule
          \textbf{Symbol} & \textbf{Value} & \textbf{Description} \\
          \hline
          $T_{\textrm{inter}}$ & 60.0~s & pedestrians inter arrival time \\
          $N_{\textrm{ped}}$  & 1..10  & number of pedestrians \\
          $B$                 & 415x394~m & simulation bound \\
          & 1.34~$\textrm{ms}^{-1}$ & mean of ped.  speed-distribution\\
          & 0.26~$\textrm{ms}^{-1}$ & std. dev. of ped. speed-distribution \\
          & 0.195~m & pedestrian radius \\
          & 0.45~m & ped. potential intimate space width \\
          & 1.2~m  & ped. potential personal space width \\
          & 50.0 & ped. potential height \\
          & 0.8 m & obstacle potential width \\
          & 6.0  & obstacle potential height \\
          & 1.2 & intimate space factor \\
          & 1  & personal space power \\
          & 1  & intimate space power \\
          \bottomrule
      \end{tabulary}
  \end{subtable}%
  \begin{subtable}{.6\linewidth}
    \centering
      \caption{Communication}
      \tiny
      \centering
      \settowidth\tymin{\textbf{Symbol}}
      \setlength\extrarowheight{3pt}
      \begin{tabulary}{0.95\linewidth}{LLL}
          \toprule
          \textbf{Symbol} & \textbf{Value} & \textbf{Description} \\
          \hline
          $T_{\textrm{beacon}}$ & 1.0 s & inter-transmission time beacon packets \\
          $S_{\textrm{beacon}}$ & 224 B & payload of beacon packets \\
          $T_{\textrm{map}}$  & 2.0 s & inter-transmission time map packets \\
          $S_{\textrm{map}}$  & 1000 B & payload of map packets \\
          \hline
          $B$ & 2.6 GHz & carrier frequency \\
          $N_{\textrm{RB}}$   & 20  & number of resource blocks \\
          $M_{\textrm{PL}}$ & URBAN-  & pathloss scenario (ITU-R M2135-1)\\
                            & MICROCELL & \\
          $h_{\textrm{UE}}$ & 1.5 m & hight of user equipment \\
          $h_{\textrm{eNB}}$ & 25.0 m & hight of eNodeB \\
          $P_{\textrm{TX, eNB}}$ & 20 dBm & transmission power of eNB (reduced) \\
          $P_{\textrm{TX, UE}}$ & 20 dBm & transmission power of UE \\
          $D_{\textrm{RLC}}$ & UM & RLC type (UM: unacknowledged mode)\\
          $Q_{\textrm{mac}}$  & 10000 B & mac queue size \\
          $Q_{\textrm{rlc\_um}}$  & $5\times 10^6$ B & rlc queue size \\
          \bottomrule
      \end{tabulary}
  \end{subtable} 
\end{table}

\label{sect:bib}
\bibliographystyle{plain}
\bibliography{Communication,CollectiveDynamics}
\end{document}